\documentstyle[preprint,aps,psfig,amsmath]{revtex}

\begin{document}

\begin{center} Degeneracies when T=0 Two Body Matrix Elements 
are Set Equal to Zero and Regge's 6j Symmetry Relations\medskip \\ 
Shadow J.Q. Robinson$^{a}$ and Larry Zamick$^{a,b}$

\noindent a) Department of Physics and Astronomy,
Rutgers University, Piscataway, \\New Jersey  08855

\noindent b)  TRIUMF,4004 Wesbrook Mall, Vancouver, British 
Columbia, \\Canada, V6T 2A3

\end{center}

\bigskip
\begin{abstract}
The effects of setting all T=0 two body interaction matrix elements
equal to a constant (or zero) in shell model calculations (designated
as $<T=0>=0$) are investigated.  Despite the apparent severity of such
a procedure, one gets fairly reasonable spectra.  We find that using
$<T=0>=0$ in single j shell calculations degeneracies appear e.g. the
$I=\frac{1}{2} ^{-}$ and $\frac{13}{2}^{-}$ states in $^{43}$Sc are at
the same excitation energies;  likewise the
I=$3_{2}^{+}$,$7_{2}^{+}$,9$^{+}_{1}$ and 10$^{+}_{1}$ states in
$^{44}$Ti.  The above degeneracies involve the vanishing of certain 6j
and 9j symbols.  The symmetry relations of Regge are used to explain
why these vanishings are not accidental.  Thus for these states the
actual deviation from degeneracy are good indicators of the effects of
the T=0 matrix elements.  A further indicator of the effects of the
T=0 interaction in an even - even nucleus is to compare the energies
of states with odd angular momentum with those that are even.

\end{abstract}
\vspace{2.5in}

\newpage

\section{Introduction}

In the early 1960's single j shell calculations in the f$_{7/2}$
region were performed by McCullen, Bayman, and Zamick (MBZ) [1,2] and
Ginocchio and French[3].  In these calculations the two body matrix
elements were taken from experiment.  However the T=0 neutron proton
spectrum in $^{42}$Sc was not well determined.  Calculations with
correct T=0 matrix elements were later performed by Kutschera, Brown,
and Ogawa[4].

In order to see how neutron-proton two body matrix elements with
isospin T=0 affect the low lying spectra of nuclei, we have set them
to a constant in a single j shell calculation in the f$_{7/2}$ region.
We can then write $V^{T=0}=c(1/4- t_{1}\cdot t_{2})$ where c is a
constant.  Hence $\sum_{i<j}V_{ij}^{T=0}=c/8(n(n-1)+6)-c/2T(T+1)$.
This means that the spectrum of states of a given isospin e.g. T=0 in
$^{44}$Ti is independent of what the constant is, it might as well be
\underline{zero}.  What the constant is will affect only the energy
splittings of states with different isospin.  We shall denote this
matrix element input as $<T=0>=0$.

Although setting all T=0 matrix elements to a constant may seem like a
severe approximation, it will be seen that one gets a fairly good
representation of the spectrum.  When the T=0 matrix elements are
reintroduced, there is some fine tuning which improves the spectrum.

While the problem of T=1 pairing is better understood and studied,
there exists a very extensive literature on the
possibility of T=0 pairing, both pro and con. We here include some of
the relevant references.[5-13]

In a shell model calculation the effects of both T=0 and T=1 pairing
are automatically included. The problem then is to sort out as much as
possible the individual effects.

In the next sections we will consider calculations in the f$_{7/2}$ shell and in the full fp space.

\section{Results of Single j shell calculations}

In the following tables we show T=T$_{min}$ calculated yrast spectra
for $^{43}$Ti(Table I) and $^{44}$Ti(Table II) where we use 2
different sets of matrix elements.  In the first two columns  
we show $<T=0>=0$ for the
$^{42}$Sc matrix elements.  The last two columns  consists 
of matrix elements from
$^{42}$Sc with the T=0 matrix elements now included.  Also to 
gain some insight into how configuration mixing
affects our results, we present full fp space
results for $^{43}$Ti and $^{44}$Ti in Tables III and IV respectively.

In the single j shell calculation for which the matrix elements were
taken from the spectrum of $^{42}$Sc the values of these matrix
elements for J =0 to J=7 were 0.000 MeV, 0.6110 MeV, 1.5863 MeV,
1.4904 MeV, 2.8153 MeV, 1.5101 MeV, 3.242 MeV, and 0.6163 MeV
respectively.  The yrast spectrum is also shown in Fig 1.  Note that
with a j$^2$ configuration the even J states have T equal to one and the odd J
T equal to zero. This is also true experimentally for these levels.  Note that the
J=1$^{+}$ and 7$^{+}$ are nearly degenerate near 0.6 MeV and the
J=3$^{+}$ and 5$^{+}$ are nearly degenerate near 1.5 MeV.  Thus the
act of setting T=0 matrix elements to a constant is equivalent to
moving the J=1$^{+}$ and 7$^{+}$ together up about 0.9 MeV.  Or
putting it another way, the act of removing the degeneracy is to lower
the energies of the J=1$^{+}$ and 7$^{+}$ by about the same
amount. This is in contrast to most studies in which only the effects
of lowering the J=1$^{+}$ state are studied.

We will point out several features to be found in the tables. We
observe many levels that were considerably separated in the 'normal'
interaction become degenerate when we go to $<T=0>=0$.  We explore
this further in the next section.  We find that in general with few
exceptions that the odd I levels of $^{44}$Ti are at a
lower excitation energy when we go to the $<T=0>=0$ version of the
interactions and that the $^{43}$Ti spectra is lowered in total.

\section{The Degeneracies that occur in $<T=0>=0$ and explanations}

As can be seen from Tables I and II some energy levels are degenerate
when the T=0 matrix elements are set equal to a constant.  The
degenerate pairs $(I_{1},I_{2})$ include

$^{43}$Ti $(\frac{1}{2}^{-},\frac{13}{2}^{-})
(\frac{17}{2}^{-},\frac{19}{2}^{-})$

$^{44}$Ti $(9^{+},10^{+})$

The wavefunctions for the Titanium isotopes are written as 
\begin{equation}
\psi = \Sigma D^{I \alpha}(J_{p},J_{n})[(j^{2})^{J_{p}}(j^{n})^{J_{n}}]^{I \alpha}
\end{equation} 
where $D^{\pm}(J_{p},J_{n})$ is the probability amplitude that in a
state of total angular momentum I the protons couple to $J_{p}$ and
the neutrons to $J_{n}$.  The elements $D^{I}(J_{p},J_{n})$ form a
column vector.

Let us first consider $(\frac{1}{2}^{-},\frac{13}{2}^{-})$ in $^{43}$Ti.  The
basis states can be written as $[J_{p},J_{n}]^{I}$ where $J_{p}$ is
the angular momentum of the two protons.  The interaction matrix
element $<[J^{\prime}_{p},J_{n}]^{I}V[J_{p},J_{n}]^{I}>
= \delta _{J_{p}^{\prime},J_{p}}E_{J_{p}}+2 \Sigma_{J} U(jjIj,J_{p}^{\prime}J)
U(jjIj,J_{p}J)E_{J}$ where $E_{J}$ is the two particle matrix element
$<[jj]^{J}V[jj]^{J}>$.  For even J, T is equal to one while for odd J,
T is equal to zero.

We next consider $^{44}$Ti
The interaction matrix element
$<[J_{p}^{\prime}J_{n}^{\prime}]^{I}V[J_{p}J_{n}]^{I}>$ is given by $E_{J_{p}}
\delta_{J_{p}J_{p}^{\prime}} \delta_{J_{n}J_{n}^{\prime}}+ E_{J_{n}}
\delta_{J_{p}J_{p}^{\prime}} \delta_{J_{n}J_{n}^{\prime}}+4
\Sigma_{JJ_{A}}<(jj)^{J_{p}^{\prime}}(jj)^{J_{n}^{\prime}}|(j)^{J}(jj)^{J_{A}}>^{I}<(jj)^{J_{p}}(jj)^{J_{n}}|(j)^{J}(jj)^{J_{A}}>^{I}
E_{J}$ where the unitary recouping coefficients are related to the
Wigner 9j symbols
\begin{equation}
<(ab)^{c}(de)^{f}|(ad)^{x}(be)^{y}>^{I}=\sqrt{(2c+1)(2f+1)(2x+1)(2y+1)}
\left\{ \begin{array}{ccc} a & b & c\\ d & e & f\\ x & y& I\\
\end{array} \right\}.
\end{equation}  For symmetry relations the 9j symbols are more convenient than the unitary coefficients. 

It is instructive to look at the energies and wavefunctions (ie column
vectors) for the $I=\frac{1}{2}^{-}$ and $I=\frac{13}{2}^{-}$ states
that appear in the NYO Technical reports (which included T=0 matrix
elements.)

\begin{tabular}{ccccc}
            &          &\hspace{.6in}$I=\frac{1}{2}$&\hspace{.6in}$I=\frac{13}{2}$  &\hspace{.6in}  \\
Energy(MeV) &          & 5.4809        & 3.8477           & 5.8122\\
$J_{p}$     &$ j_{n}$  &               &                  & \\
4           & 7/2      &1.000          & 0.9942           & -0.1076\\
6           & 7/2      &0.000          & 0.1076           & 0.9942\\
\end{tabular}

In the $f_{7/2}$ model the $I=\frac{1}{2}^{-}$ configuration is unique
$[J_{p}=4$ $ J_{n}=\frac{7}{2}]^{\frac{1}{2} -}$. There are two configurations
for the $I=\frac{13}{2}^{-}$ state $[4 \frac{7}{2}]$ and $[6
\frac{7}{2}]$.

When we go to $<T=0>=0$ what basically happens is that the eigenvalues
and eigenfunctions become

\begin{tabular}{ccccc}
            &          &\hspace{.6in}$I=\frac{1}{2}$&\hspace{.6in}$I=\frac{13}{2}$  &\hspace{.6in}  \\
            &          &$E_{1}$        & $E_{1}$          & $E_{2}$\\
$J_{p}$     &$ j_{n}$  &               &                  &      \\
4           & 7/2      &1.000          & 1.000            & 0.000\\
6           & 7/2      &0.000          & 0.000            & 1.000\\
\end{tabular}

In order for this to happen the matrix element
$<[J_{p}=4,j_{n}=\frac{7}{2}]^{I=\frac{13}{2}}V[J_{p}=6,j_{n}=\frac{7}{2}]^{I=\frac{13}{2}}>$
must vanish. This vanishing is carried by the Racah coefficients
$U(\frac{7}{2}\frac{7}{2}\frac{13}{2}\frac{7}{2};4J)U(\frac{7}{2}\frac{7}{2}\frac{13}{2}\frac{7}{2};6J)$
where J is the angular momentum of a neutron-proton pair.

In general J can be 4,5,6, or 7.  However in $<T=0>=0$, only the even J's
contribute i.e. J=4 or J=6. In either case one of the Racah coefficients will
be $U(\frac{7}{2}\frac{7}{2}\frac{13}{2}\frac{7}{2};46)$.  This Racah
coefficient is zero.  This guarantees a decoupling of $[4
\frac{7}{2}]$ from $[6 \frac{7}{2}]$ but does not in itself lead to a
degeneracy of the $I=\frac{1}{2}^-$ and $I=\frac{13}{2}^-$ states. That
happens because of this additional condition

\begin{equation}
 U(\frac{7}{2} \frac{7}{2} \frac{13}{2} \frac{7}{2};44) = U
 (\frac{7}{2}\frac{7}{2}\frac{1}{2}\frac{7}{2};44)=\frac{1}{2}
\end{equation}

We next consider the degeneracy of $I=9^{+}$ and $10^{+}$ in $^{44}$Ti in
$<T=0>=0$.  It is again instructive to write down the eigenfunctuions as
they appear in the NYO report
\begin{tabular}{cccccccc}
            &          &\hspace{.6in}$I=9$&\hspace{.6in} &\hspace{.6in}         &\hspace{.6in}$I=10$  &\hspace{.6in}  \\
Energy      &          &8.7799 & 8.8590 & 11.5951 & 7.8429 & 9.8814 & 10.5110\\
Isospin     &          &       & T=1    & T=1     &        &        & T=1    \\
$J_{p}$     &$ j_{n}$  &       &        &         &        &        &        \\
4           & 6        &-0.7071&0.5636  &-0.4270  & 0.7037 &-0.0696 & 0.7071 \\
6           & 4        &0.7071 &0.5636  &-0.4270  & 0.7037 &-0.0696 &-0.7071 \\
6           & 6        &0.0000 &0.6039  & 0.7971  & 0.0984 & 0.9951 & 0.0000 \\
\end{tabular}

Before proceeding, we remind the reader of a general rule that can
clearly be seen in the wave functions above.  For even total angular
momentum I the wave functions of even T states of N=Z nuclei do not
change sign under the interchange of neutrons and protons but the T=1
wavefunctions do change sign.  For odd I it is the opposite.  This can be
summarized by $D^{IT}(J_{p},J_{n})=(-1)^{I+T}D^{IT}(J_{n},J_{p})$.

 We focus on the T=0 states.  This makes the life much
simpler.  Instead of three states each we need only worry about one
I=9$^+$ and two I=10$^+$ states.  Note that for I=9$^+$ T=0 the state was the
simple wavefunction $\left( \begin{array}{c} \frac{-1}{\sqrt{2}}\\
\frac{1}{\sqrt{2}}\\ 0\\
\end{array} \right)$.

What clearly happens for I=10$^+$ in $<T=0>=0$ is that there is a
decoupling of [6,4] and [4,6] from [6,6] So that the wavefunctions of
the two T=0 states become $\left( \begin{array}{c}
\frac{1}{\sqrt{2}}\\ \frac{1}{\sqrt{2}}\\ 0
\end{array} \right)$
and
$\left( \begin{array}{c}
0\\
0\\
1
\end{array} \right)$
and the eigenvalues of the first one becomes the same as that of the
unique I=9 state.

We further note that aside from the yrast degeneracies there are other
degenerices. For example, the $7_{2}^{+}$ and $3_{2}^{+}$ are
degenerate with the I=$9_{1}^+$,$10_{1}^+$ pair in $^{44}$Ti.  At first
this is puzzling because the dimensions are different.  There are
seven basis states for I=3$^+$ and six for I=7$^+$, whereas for I=9$^+$ and 10$^+$
there are only three basis states.  However, of the seven I=3$^+$ states,
five have isospin \underline{one}, and only two have isospin T=0.  Of
the six I=7$^+$ states, four have isospin one and only two have isospin
zero.  Since we are focusing on T=0 we only show only these
wavefunctions in Table V.  When the T=0 two particle matrix
elements are set equal to zero the wave functions simplify as shown in
the table.

We now begin to see a connection between
$I=3_{2}^{+},7_{2}^{+},9_{1}^{+},$ and $10_{1}^{+}$.  For the $9_{1}^{+}$ and
$10_{1}^{+}$ the only non-zero components of the wave function in the
$<T=0>=0$ are D(4,6) and D(6,4) both having magnitude
$\frac{1}{\sqrt{2}}$.  The $3^{+}_{1}$ state has nonzero components
D(2,4) and D(4,2).  There is no connection with the $9_{1}^{+}$ and
$10_{1}^{+}$ states.  However for the $3_{2}^{+}$ state the only
non-vanishing matrix elements are D(4,6) and D(6,4) each with magnitude
$\frac{1}{\sqrt{2}}$.  This is the same as what occurs for the
$9_{1}^{+}$ and $10_{1}^{+}$ states.

A similar story is written by I=7$^+$.  The non vanishing components for the
$7_{1}^{+}$ state in the $<T=0>=0$ case are D(2,6) and D(6,2) however
for the $7_{2}^{+}$ state they are D(4,6) and D(6,4) each with magnitude
$\frac{1}{\sqrt{2}}$.

Thus a common theme emerges for
$I=3_{2}^{+},7_{2}^{+},9_{1}^{+},$ and $10_{1}^{+}$ (all T=0) in that for the
$<T=0>=0$ case the only non-vanishing components of the wave functions
are D(4,6) and D(6,4).  Visually, the column vectors look the same.
And it is precisely these states that are degenerate.

Let us now show why in the case of $<T=0>=0$ the matrix element
$<[J_{p}^{\prime}=4J_{N}^{\prime}=6]^{I=10}V[J_{p}=6J_{N}=6]^{I=10}>$
vanishes.  This is a necessary condition for the wave functions to
have the simple form discussed in this section.

From the expression for the neutron-proton interaction previously
given the above matrix element is (j=$\frac{7}{2}$)
\begin{equation}
 (c)(13)(9) 
\left\{ \begin{array}{ccc}
j & j& 4\\
j&j& 6\\
4 &6&10\\
\end{array} \right\} 
\left\{ \begin{array}{ccc}
j & j& 6\\
j&j& 6\\
4 &6&10\\
\end{array} \right\} E^{4}
+(c)(13) \Sigma_{J_A} (2J_{A}+1)
\left\{ \begin{array}{ccc}
j & j& 4\\
j&j& 6\\
6&J_{A}&10\\
\end{array} \right\}
\left\{ \begin{array}{ccc}
j & j& 6\\
j&j& 6\\
6&J_{A} &10\\
\end{array} \right\} E^{6}
\end{equation}
where the proportionality constant c is 156$\sqrt{13}$.
(Note that $E^{5}$ and $E^{7}$ are equal to zero because all odd J
have T=0) Because the last 9j above has two rows identical it is
necessary for $J_{A}$ to be even ie $J_{A}$=4 or 6.  Thus the coefficient
of $E^{6}$ is
\begin{equation} 
(c)(13)(9)
\left\{ \begin{array}{ccc}
j & j& 4\\
j&j& 6\\
6 &4&10\\
\end{array} \right\}
\left\{ \begin{array}{ccc}
j & j& 6\\
j&j& 6\\
6 &4&10\\
\end{array} \right\}
+(c)(13)(13)
\left\{ \begin{array}{ccc}
j & j& 4\\
j&j& 6\\
6 &6&10\\
\end{array} \right\}
\left\{ \begin{array}{ccc}
j & j& 6\\
j&j& 6\\
6&6&10\\
\end{array} \right\}
\end{equation}

Using symmetry properties of 9j symbols we note that every term in the
above expression (both for $E^{4}$ and $E^{6}$) contains the 9j symbol $
\left\{ \begin{array}{ccc} j & j& 6\\ j&j& 6\\ 6 &4&10\\
\end{array} \right\}
$. This 9j symbol is zero and hence we have shown why the above
neutron-proton matrix element vanishes. It is by no means  
obvious why this 9j vanishes. There will be considerable discussion in the
next section of why some of the 6j's and 9j's we encounter vanish.

Although in Table V we have only shown T=0 wave functions there are
several T=1 states interspaced amongst the T=0 states.  For example,
in the Technical Report NYO-9891 [2] for I=3$^+$ the lowest state
calculated to be at 6.2357 MeV has T=1.  The calculated energy for
this state is about 300 keV lower than the lowest T=0 state shown in
Table V.  Other T=1 states are calculated to be at 9.2334, 10.0321
and 10.9022 MeV.  For I=7$^+$ the lowest T=1 state is calculated to be at
6.7094 MeV, just above the other the lowest T=0 state shown in Table
V.  The other T=1 states for I=7$^+$ are calculated to be at 9.0744,
9.5141 and 12.1535 MeV.  The closeness of T=0 and T=1 states was
previously discussed by Goode and Zamick [15].

\section{Why some Racah coefficients vanish - Regge Symmetries}

Thus far we have explained how degeneracies arise by matrices that
certain Racah or 9j symbols vanish.  In this section we look for a
deeper meaning.  We were aided in this by many insightful articles 
collected in Biedenharn and Van Dam[16].

For convenience we shall switch from unitary Racah coeffiecients to Wigner 6j symbols
\begin{equation}
U(abcd;ef)=(-1)^{a+b+c+d} \sqrt{(2e+1)(2f+1)} \left\{ \begin{array}{ccc}
a & b & c\\
d&e& f\\
\end{array} \right\}
\nonumber
\end{equation}

In the previous section we noted that the 6j symbol 
$ \left\{ \begin{array}{ccc}
\frac{7}{2} & \frac{7}{2} & 4\\
\frac{7}{2}&\frac{13}{2}& 6\\
\end{array} \right\}$ vanished.  We note that this is a particular case of a 
wider class of 6j's that vanish.  All 6j's of the form 
$ \left\{ \begin{array}{ccc}
j & j & (2j-3)\\
j&(3j-4)& (2j-1)\\
\end{array} \right\}
$ vanish for all j, both half integer and integer.  Besides the six j above other examples are $
 \left\{ \begin{array}{ccc}
\frac{5}{2} & \frac{5}{2} & 2\\
\frac{5}{2}&\frac{7}{2}& 4\\
\end{array} \right\}
,
 \left\{ \begin{array}{ccc}
\frac{9}{2} & \frac{9}{2} & 6\\
\frac{9}{2}&\frac{19}{2}& 8\\
\end{array} \right\} 
$, and $
 \left\{ \begin{array}{ccc}
4 & 4 & 5\\
4&8& 7\\
\end{array} \right\}
$.

We find we can relate the above 6j symbol to a simpler one using one of
the six remarkable relations discovered by Regge in 1959
[17] We follow the notation of Rotenberg et. al. [18]\\
$\left\{ \begin{array}{ccc}
j_{1} & j_{2} & j_{3}\\
l_{1}&l_{2}& l_{3}\\
\end{array} \right\}
=
 \left\{ \begin{array}{ccc}
A & B & C\\
D&E& F\\
\end{array} \right\}
$\\
A=$\frac{1}{2}(j_1+j_2+l_1-l_2)$ \hspace{1in} B=$\frac{1}{2}(j_2+j_3+l_2-l_3)$\\
C=$\frac{1}{2}(j_1+j_3-l_1+l_3)$ \hspace{1in} D=$\frac{1}{2}(j_1-j_2+l_1+l_2)$\\
E=$\frac{1}{2}(j_2-j_3+l_2+l_3)$ \hspace{1in} F=$\frac{1}{2}(-j_1+j_3+l_1+l_3)$\\

From this Regge symmetry relation we find that 
\begin{equation}
\left\{ \begin{array}{ccc}
j & j & (2j-3)\\
j& (3j-4)&(2j-1)\\
\end{array} \right\}
=
\left\{ \begin{array}{ccc}
2 & (2j-3) & (2j-2)\\
(2j-2)&(2j-1)& (2j-2)\\
\end{array} \right\}
=
\left\{ \begin{array}{ccc}
(2j-2) &(2j-3) & 2\\
(2j-2) &(2j-1) &(2j-2)\\
\end{array} \right\}
\end{equation}

We note that 6j symbols with a ``two'' in them have been worked out by
Biedenharn, Blatt, and Rose [19].  Using their notation we find from
their results that 
$ \left\{ \begin{array}{ccc} l_{1} & J_{1} & 2\\
J_{2}&l_{2}& L\\
\end{array} \right\}
$
 for $l_{2}=J_{1}+1$ and $l_{1}=J_{1}+1$ is proportional to X where 
\begin{equation}
X=[(J_1+1)(J_1-J_2)-L(L+1)+J_2(J_2+2)]
\end{equation}

We have $L=2j-2$, $J_1=2j-3$, $l_1=2j-2$, $J_2=2j-2$,
and $l_2=2j-1$.  With these values we see that X vanishes.

In Regge's paper [17] he states ``although no direct connection has
been established between these wider symmetries it seems very probably
that it will be found in the future.''  He also states ``We see
therefore that there are 144 identical Racah's coefficients.... It
should be pointed out that this wider 144-group is isomorphic to the
direct product of the permutation group of 3 and 4 objects.''

Following Regge's work Bargmann presented, amongst other things, his
derivation of the Regge symmetries [20].  He there stated ``While the
following analysis does not lead to a deeper understanding of the
Regge symmetries it yields, at least a fairly transparent derivation
of the symmetries.''

In section III we pointed out that a certain 9j symbol
``unexpectedly'' vanished.  Perhaps there are some symmetries involving
the 9j symbols as well.  The only comment by Bargmann on this [20] is
``Schwinger has computed the generating function for the 9j symbol.
This does not reveal any new symmetries - at least none to be obtained
by a permutation of the relevant quantities k$_{\alpha \beta}$.''

%begin new stuff

Nevertheless the Regge symmetries for 6j symbols do have some
implications for 9j's.  The 9j mentioned in the previous section
$\left\{ \begin{array}{ccc}
\frac{7}{2} &\frac{7}{2} & 6\\
\frac{7}{2} &\frac{7}{2} & 6\\
     4      &     6      &10\\
\end{array} \right\}
$
is part of a wider class of identically zero 9j symbols. These
are of the form
$
\left\{ \begin{array}{ccc}
 j     & j      & (2j-1)\\
 j     & j      & (2j-1)\\
 (2j-1)& (2j-3) & (4j-4)\\
\end{array} \right\}
$. ( Other examples
are
$\left\{ \begin{array}{ccc}
\frac{9}{2} &\frac{9}{2} & 8\\
\frac{9}{2} &\frac{9}{2} & 8\\
     8      &     6      &14\\
\end{array} \right\}
$
and
$
\left\{ \begin{array}{ccc}
4 &4 & 7\\
4 &4 & 7\\
7 &5 &12\\
\end{array} \right\}
$
.)

Following the notation of Rotenberg et.al.[18] we first
write down the well known expression for a 9j as a sum over three 6j
symbols.
\begin{eqnarray}
\left\{ \begin{array}{ccc}
 j     & j      & (2j-1)\\
 j     & j      & (2j-1)\\
 (2j-1)& (2j-3) & (4j-4)\\
\end{array} \right\}
= 
\Sigma_{\beta} (-1)^{2 \beta} (2 \beta +1) \nonumber \\
\left\{ \begin{array}{ccc}
 j     & j      & (2j-1)\\
 (2j-3)     & (4j-4)      & \beta \\
\end{array} \right\}
\left\{ \begin{array}{ccc}
 j     & j      & (2j-3)\\
 j     & \beta      & (2j-1)\\
\end{array} \right\}
\left\{ \begin{array}{ccc}
 (2j-1)     & (2j-1)      & (4j-4)\\
 \beta     & j      & j             \\
\end{array} \right\}
\end{eqnarray}

The parameter $\beta$ is constrained by triangle 
relations in each of the 6j symbols.  In particular
first 6j symbol constrains $\beta$ as follows
\begin{eqnarray}
\beta \ge (3j-4) \\
\beta \le (3j-3).
\end{eqnarray}

From these constraints $\beta$ = (3j-3) or 
(3j-4) or the first 6j symbol is zero.  If 
$\beta$ = (3j-4) the second 6j symbol becomes the 
one previously discussed above in equation 6 and was there
shown to be zero.  This leaves $\beta$ = (3j-3).

In this case the last 6j symbol becomes 
$
\left\{ \begin{array}{ccc}
 (2j-1)     & (2j-1)      & (4j-4)\\
 (3j-3)     & j      & j             \\
\end{array} \right\}
$ which we now show vanishes.  

We will use the Regge symmetry [18] 
\\
$\left\{ \begin{array}{ccc}
j_{1} & j_{2} & j_{3}\\
l_{1}&l_{2}& l_{3}\\
\end{array} \right\}
=
 \left\{ \begin{array}{ccc}
A & B & C\\
D&E& F\\
\end{array} \right\}
$\\
$
\begin{array}{cc}
A=j_1 & \hspace{0.5in} B=\frac{1}{2}(j_2+j_3-l_2+l_3)\\
C=\frac{1}{2}(j_2+j_3+l_2-l_3) & D=l_1\\
E=\frac{1}{2}(-j_2+j_3+l_2+l_3) & \hspace{0.5in} F=\frac{1}{2}(j_2-j_3+l_2+l_3)\\
\end{array}
$
\\
so that we can now write
\begin{equation}
\left\{ \begin{array}{ccc}
 (2j-1)     & (2j-1)      & (4j-4)\\
 (3j-3)     & j      & j          \\
\end{array} \right\}
=
\left\{ \begin{array}{ccc}
 (2j-1)     & (3j-\frac{5}{2})      & (3j-\frac{5}{2})\\
 (3j-3)     & (2j-\frac{3}{2})      & \frac{3}{2}\\
\end{array} \right\}
=
\left\{ \begin{array}{ccc}
  (3j-\frac{5}{2})    & (3j-\frac{5}{2})      & (2j-1)\\
 \frac{3}{2}     &     (2j-\frac{3}{2})   & (3j-3)\\
\end{array} \right\}
\end{equation}

The results of 6j symbols with a ``$\frac{3}{2}$'' are 
found in Varshalovich, Moskalev, and Khersonski [21]
$
\left\{ \begin{array}{ccc}
 a     & b      & c\\
 \frac{3}{2}     & e      & f\\
\end{array} \right\}
$ for e=c-$\frac{1}{2}$ and f=b-$\frac{1}{2}$, 
as we have here, is proportional to
\begin{equation}
3(-a(a+1)+b(b+1)+c(c+1))-2(b+1)(c+1)
\end{equation}
which for $a=(3j-\frac{5}{2})$, $b=(3j-\frac{5}{2})$, 
and $c= (2j-1)$ is zero.  Thus in the lone remaining 
case of $\beta$=(3j-3) the final 6j symbol in the sum 
is zero.  So for any allowed value of $\beta$ one of the 6j 
symbols is zero implying that the 9j symbol 
$
\left\{ \begin{array}{ccc}
 j     & j      & (2j-1)\\
 j     & j      & (2j-1)\\
 (2j-1)& (2j-3) & (4j-4)\\
\end{array} \right\}
$ is zero.

%end new stuff

\section{Full f-p calculation for $^{43}Ti$ and $^{44}Ti$}

We have performed full fp calculations for $^{44}$Ti and $^{43}$Ti
with the FPD6 interaction.  We shall show these and also compare the
$^{44}$Ti calculations with single j results using the spectrum of
$^{42}$Sc as input.  The later is shown in Fig 1.

We first discuss T=0 states in the even-even nucleus $^{44}$Ti.  In
Table II and Fig 2 we show the single j results.  The first two
columns show the results when the T=0 two body matrix elements are set
to zero i.e. $<T=0>=0$.  In Fig 2 we show the even I states of $^{44}$Ti in the
first column and the odd I in the second column.  Note that the I=9$^+$
and I=10$^+$ states are degenerate as has been previously discussed.

In the last two columns we have the single j shell results when the
full spectrum of $^{42}$Sc is introduced including the T=0 matrix
elements.  We note that there is much more change in the odd I
spectrum than in the even I.  The odd I spectrum raises considerably.
The even I spectrum gets spread out a bit but this is tame in
comparison to the alteration in the odd I spectrum.

In Table IV and Fig 3 we show results for a full f-p calculation using FPD6.
We use the same format as for Table II.  When the two body T=0 matrix
elements are set equal to zero (first two columns), we find
surprisingly that there is not much difference with the single j shell
calculation shown in Table II and Fig 2.  The I=9$^+$ and 10$^+$ state which were
exactly degenerate in the single j shell calculation are still nearly
degenerate in the full fp calculation.  The overall spectra do not
look very different (see first two columns in Tables II and IV and Fig 2 and 3).

There is one difference however the appearance in Table IV and Fig 3 of $I=1^+$
and $11^+$ T=0 states.  In a single j shell calculation the $I=1^+$
and $11^+$ states all have isospin T=1.

We now come to the full f-p calculation in which all the two body
matrix elements of the FPD6 interaction are in play- both T=0 and
T=1.  Now we see major differences for both the even I and odd I
states of $^{44}$Ti. (See Table IV and Fig 3 right hand columns).

If we look at the low spin states, $I=0^+,2^+$, and $4^+$ they are
largely unaffected when the T=0 two body matrix elements are put back
in.  The main difference comes from the higher spin states.  With the
full FPD6 the spectrum of the even I gets spread out more looking
somewhat rotational.  For example the I=10$^+$ state increases in energy
from 6.476 MeV to 7.790 MeV.  In the corresponding single j shell
calculation there was hardly any change in the I=10$^+$ energy. Likewise the
I=12$^+$ energy goes up from 7.192 MeV to 8.574 MeV when the
T=0 two body matrix elements are put back into FPD6.

The odd I states experience a substantial upward shift in the
spectrum.  Now the I=9$^+$ state is considerably higher than the $I=10^+$
state (9.030 vs 7.790 MeV).

In the single j shell calculation with matrix elements from $^{42}$Sc
 the even I columns corresponding to $<T=0>=0$ and full spectrum (the first and 
third columns of energy levels) are not that different.  It appears
that the reintroduction of the T=0 two body matrix elements does not make much difference.
In Fig 3 however the third column, again even I, gets more spread out
 relative to the first column going a bit in the direction of giving 
a more rotational spectrum.  Thus is would appear that for even I the 
T=0 two body matrix elements will affect the spectrum in a significant 
way only when configuration mixing is present.

We now consider the odd-even spectrum $^{43}$Ti($^{43}$Sc).  The
results are shown in Tables I and III and in Fig 4.  In the figure we
only show a full calculations with FPD6 and compare results when the
T=0 two body matrix elements are set equal to zero (first column) with
those where the full FPD6 interaction is included (second column).

The results at first look a bit complicated but a careful examination
shows systematic behavior.

For I less than $\frac{7}{2} ^-$ the states come down in energy
(relative to the I=$\frac{7}{2}^-$ ground state).  For I greater than
$\frac{7}{2}^-$ there is another systematic.  When the T=0 two body
matrix elements are set to zero there are nearly degenerate doublets
($\frac{9}{2}^-$,$\frac{11}{2}^-$) ($\frac{13}{2}^-$,$\frac{15}{2}^-$)
and ($\frac{17}{2}^-$,$\frac{19}{2}^-$).  The effect of putting T=0 two
body matrix elements back in is to cause the lower spin member of each
doublet to rise in energy by a substantial amount, while the higher
spin member lowers itself a small amount, ie I=
$\frac{9}{2}^-$,$\frac{13}{2}^-$, and $\frac{17}{2}^-$ rise noticeably
but I= $\frac{11}{2}^-$,$\frac{15}{2}^-$, and $\frac{19}{2}^-$ drop
slightly.  This spectral staggering should be good evidence of the
importance of T=0 two body matrix elements.

Work on the effect of L=0, T=1 and L=1, T=0 pairing in the f-p shell
has already been performed by Poves and Martinez-Pinedo.[22] They start
with a realistic interaction, KB3, and study the effects of removing
the T=1 pairing from the T=0 S=1 pairing.  They focused on binding
energies and on the even spin states of $^{48}$Cr.  Relative to their
work, whose conclusions we certainly agree with, we have made a more
severe approximation of setting all T=0 matrix elements equal to zero.
The payoff for us is that certain degeneracies appear between states,
the deviation of which in the physical spectrum can largely be
attributed to T=0 two body matrix elements.  Also, we focussed on odd
I excited states.  The deviation in the physical spectrum of the
energies of odd I states from even I is also a good indication of the
effects of T=0 matrix elements.

\bigskip

This work was supported by the U.S. Dept. of Energy under Grant
No. DE-FG02-95ER-40940 and one of us by a GK-12 NSF9979491
Fellowship(SJQR).  One of us (L.Z.) is grateful for the hospitality
afforded to him at the Institute for Nuclear Theory in Seattle where many of
these ideas were crystallized.

\begin{table}
\caption{Spectra of $^{43}$Ti}
\begin{tabular}{cccc}
$^{42}$Sc $<T=0>=0$ interaction &     & $^{42}$Sc interaction&      \\
I   &  E(MeV) &     I   &E(MeV)\\
7/2 & 0.0000  &7/2      &0.000 \\  
9/2 & 1.640   &9/2      &1.680 \\
3/2 & 1.831   &11/2     &2.335 \\
11/2& 2.061   &3/2      &2.888 \\
5/2 & 2.832   &5/2      &3.449 \\
1/2 & 3.279   &13/2     &3.500 \\
13/2& 3.279   &15/2     &3.511 \\
15/2& 3.425   &19/2     &3.644 \\
17/2& 3.919   &17/2     &4.298 \\
19/2& 3.919   &1/2      &4.316 \\
\end{tabular}
\end{table}

\begin{table}
\caption{Spectra of $^{44}$Ti}
\begin{tabular}{cccc}
$^{42}$Sc $<T=0>=0$ interaction &     & $^{42}$Sc interaction  &      \\
I   &  E(MeV) &     I  &E(MeV) \\
0   & 0.000   &0       &0.000  \\     
2   & 1.303   &2       &1.163  \\
4   & 2.741   &4       &2.790  \\
6   & 3.500   &6       &4.062  \\
3   & 4.716   &3       &5.786  \\
5   & 4.998   &5       &5.871  \\
7   & 5.356   &7       &6.043  \\
8   & 5.656   &8       &6.084  \\
9   & 7.200   &10      &7.384  \\
10  & 7.200   &12      &7.702  \\
12  & 7.840   &9       &7.984  \\
\end{tabular}
\end{table} 

\begin{table}
\caption{$^{43}$Ti full fp calculation}
\begin{tabular}{cccc}
FPD6 $<T=0>=0$ & & FPD6  & \\
\tableline
I     &       E(MeV) &     I  &     E(MeV) \\
7/2   & 0.000        &7/2     &0.000  \\     
3/2   & 1.668        &3/2     &0.871  \\
9/2   & 1.970        &1/2     &1.805  \\
11/2  & 2.000        &11/2    &1.889  \\
5/2   & 2.638        &5/2     &2.305  \\
1/2   & 2.940        &9/2     &2.633  \\
15/2  & 3.065        &15/2    &2.948  \\
13/2  & 3.070        &19/2    &3.401  \\
17/2  & 3.325        &13/2    &3.718  \\
19/2  & 3.417        &17/2    &4.429  \\
\end{tabular}
\end{table}

\begin{table}
\caption{$^{44}$Ti full fp calculation}
\begin{tabular}{cccc}
FPD6 $<T=0>=0$   & & FPD6 & \\
\tableline
I   &       E(MeV) &     I  &     E(MeV) \\
0   & 0.000        &0       &0.000  \\     
2   & 1.515        &2       &1.317  \\
4   & 2.587        &4       &2.536  \\
6   & 3.223        &6       &3.843  \\
3   & 4.717        &3       &6.241  \\
5   & 4.932        &8       &6.383  \\
8   & 5.292        &5       &7.579  \\
7   & 5.391        &10      &7.790  \\
10  & 6.476        &7       &7.921  \\
9   & 6.574        &12      &8.574  \\
1   & 7.070        &9       &9.030  \\
12  & 7.192        &1       &9.681  \\
11  & 9.914        &11      &11.028 \\
\end{tabular}
\end{table}

\begin{table}
\caption{Comparison of wave functions of MBZ$^{a}$ with those for which $<T=0>=0$ matrix elements are set equal to zero. }
\begin{tabular}{cccccc}
I=3         &       &  MBZ   & $<T=0>=0$          & MBZ     & $<T=0>=0$                 \\
Energy(MeV) &       & 6.533  &                    & 10.493  &                           \\
$J_{P}$     &$J_{N}$&        &                    &         &                           \\
2           & 2     & 0.0000 & 0                  &0.0000   & 0                         \\
2           & 4     & 0.6968 &$\frac{1}{\sqrt{2}}$  &-0.1202  & 0                         \\
4           & 2     &-0.6968 &$\frac{-1}{\sqrt{2}}$ &0.1202   & 0                         \\
4           & 4     & 0.0000 & 0                  &0.0000   & 0                         \\
4           & 6     & 0.1202 & 0                  &0.6968   &$\frac{1}{\sqrt{2}}  $       \\
6           & 4     &-0.1202 & 0                  &-0.6968  &$\frac{-1}{\sqrt{2}}$        \\
6           & 6     & 0.0000 & 0                  &0.0000   & 0                         \\
I=7         &       &  MBZ   & $<T=0>=0$          & MBZ     & $<T=0>=0$                 \\
Energy(MeV) &       & 6.5723 &                    & 9.6570  &                           \\
$J_{P}$     &$J_{N}$&        &                    &         &                           \\
2           & 6     & 0.6965 &$\frac{1}{\sqrt{2}}$  &0.1220   & 0                         \\
4           & 4     & 0.0000 & 0                  &0.0000   & 0                         \\
4           & 6     & 0.1220 & 0                  &-0.6965  &$\frac{-1}{\sqrt{2}}$        \\
6           & 2     &-0.6965 &$\frac{-1}{\sqrt{2}}$ &-0.1220  & 0                         \\
6           & 4     &-0.1220 & 0                  &0.6965   &$\frac{1}{\sqrt{2}}$         \\
6           & 6     & 0.0000 & 0                  &0.0000   & 0                         \\
I=9         &       &  MBZ   & $<T=0>=0$          & MBZ     & $<T=0>=0$                 \\
Energy(MeV) &       & 8.7799 &                    &         &                           \\
$J_{P}$     &$J_{N}$&        &                    &         &                           \\
4           & 6     &-0.7071 &$\frac{-1}{\sqrt{2}}$&        &                           \\
6           & 4     & 0.7071 &$\frac{1}{\sqrt{2}}$&         &                           \\
6           & 6     & 0.0000 & 0                  &         &                           \\
I=10        &       &  MBZ   & $<T=0>=0$          & MBZ     & $<T=0>=0$                 \\
Energy(MeV) &       & 7.8429 &                    & 9.8814  &                           \\
$J_{P}$     &$J_{N}$&        &                    &         &                           \\
4           & 6     & 0.7037 &$\frac{1}{\sqrt{2}}$&-0.0696  & 0                         \\
6           & 4     & 0.7037 &$\frac{1}{\sqrt{2}}$&-0.0696  & 0                         \\
6           & 6     & 0.0084 & 0                  &0.9951   & 1                         \\
\end{tabular}
$a$) From Technical Report NYO 9801 [2]
\end{table} 

\begin{figure}
\psfig{file=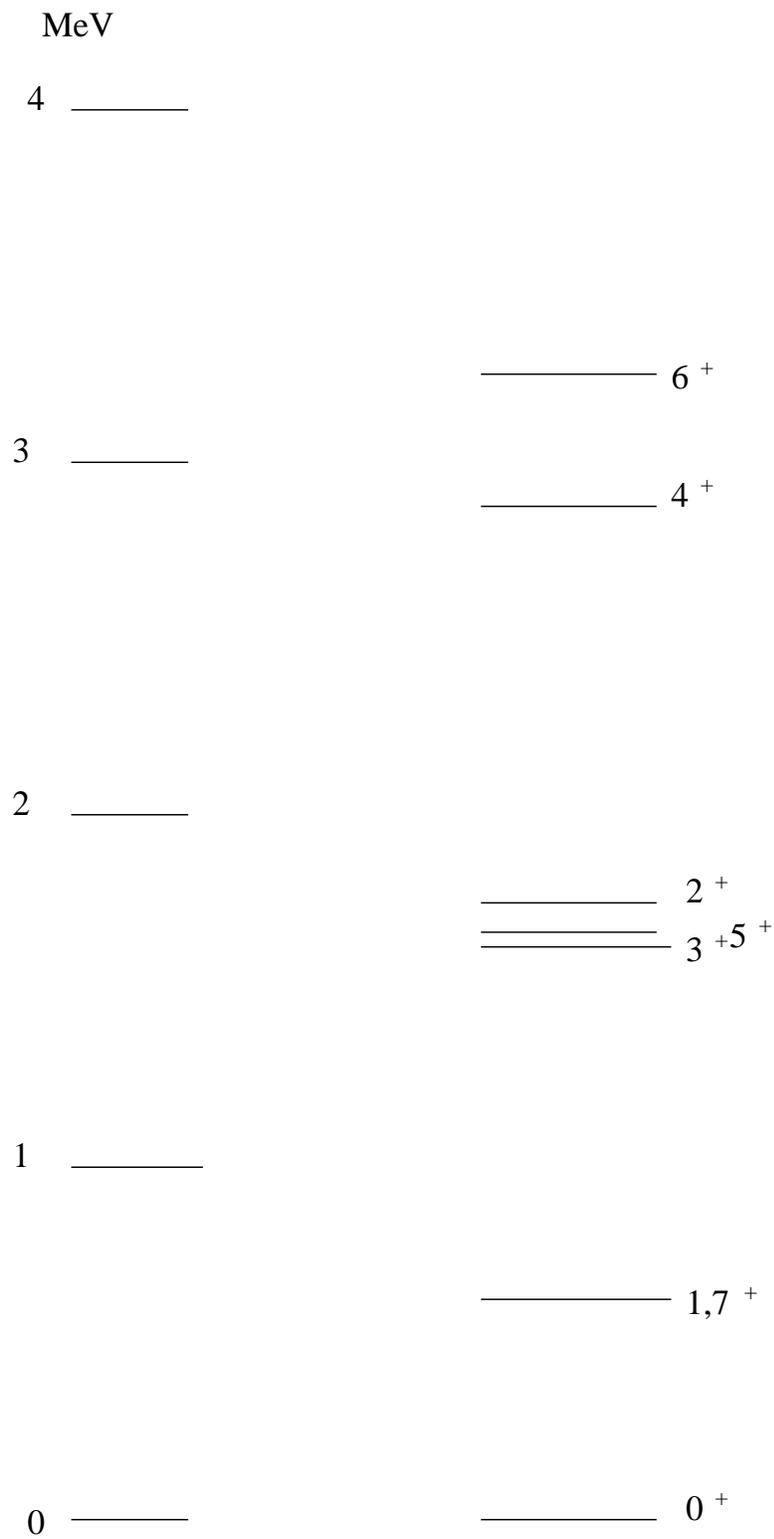,height=8in}
\caption{Spectrum of $^{42}$Sc}
\end{figure}

\begin{figure}
\psfig{file=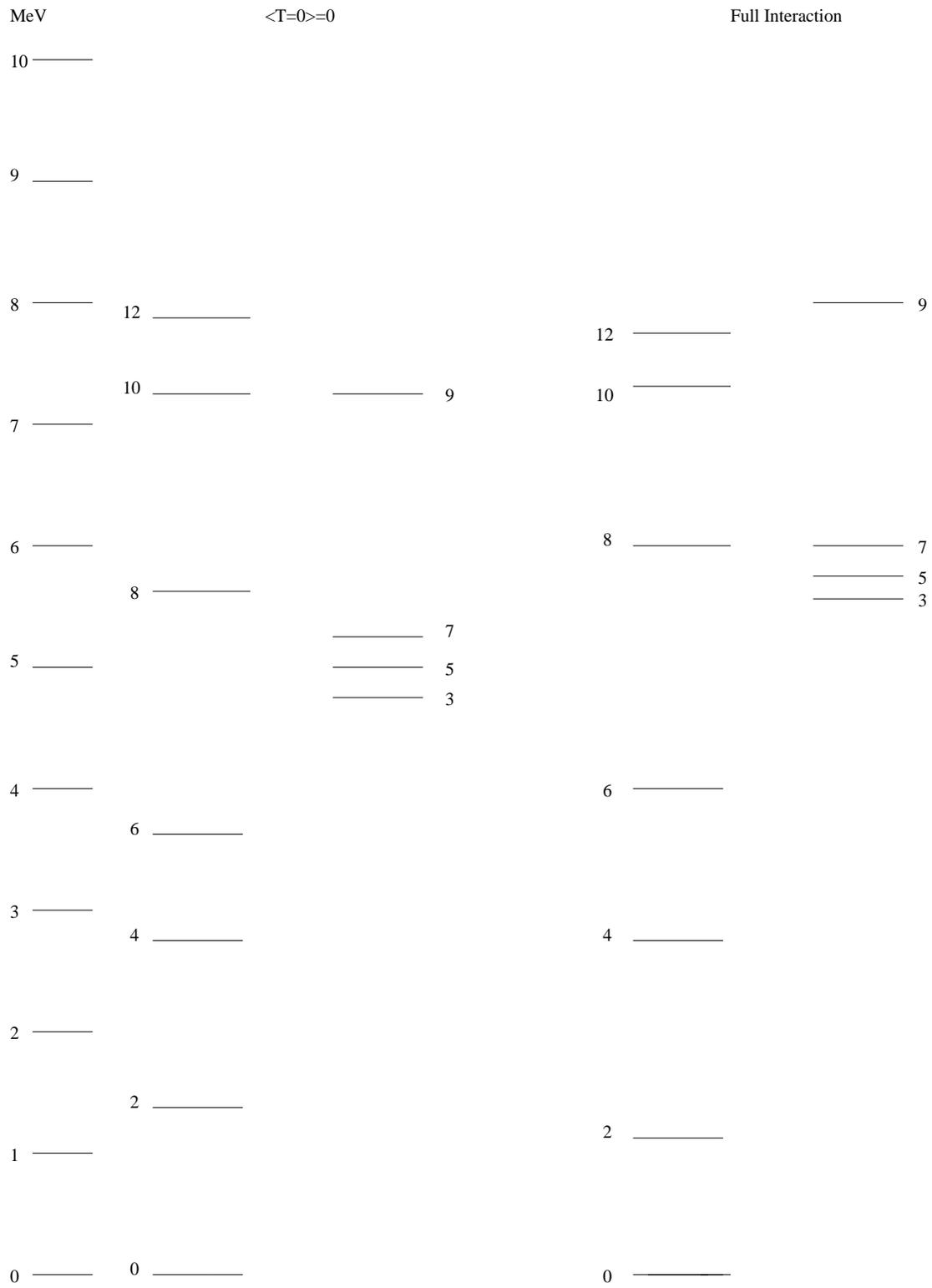,height=8in}
\caption{Single j T=0 $^{44}$Ti with matrix elements from $^{42}$Sc}
\end{figure}

\begin{figure}
\psfig{file=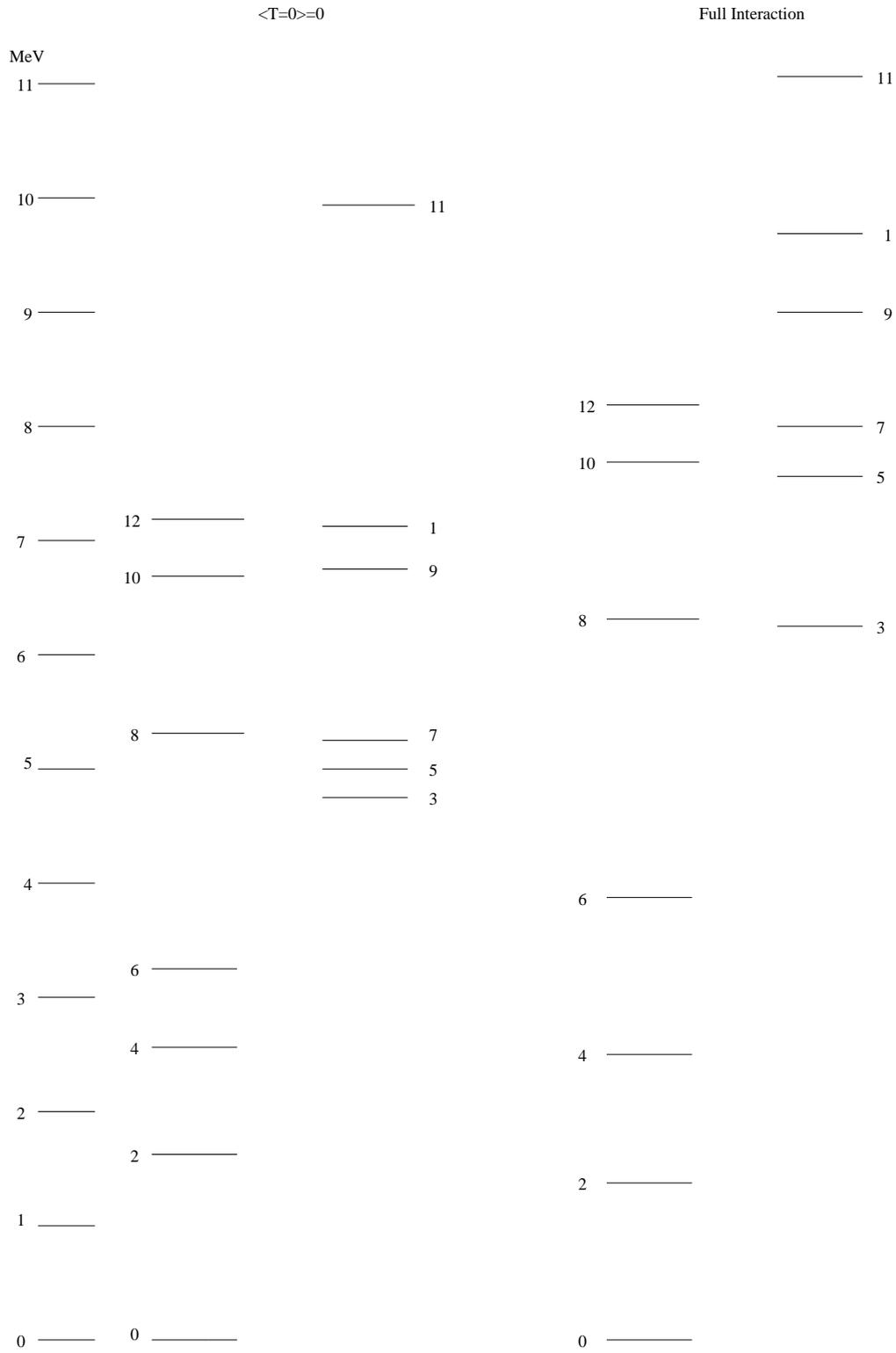,height=8in}
\caption{Full f-p T=0 $^{44}$Ti with FPD6 interaction}
\end{figure}

\begin{figure}
\psfig{file=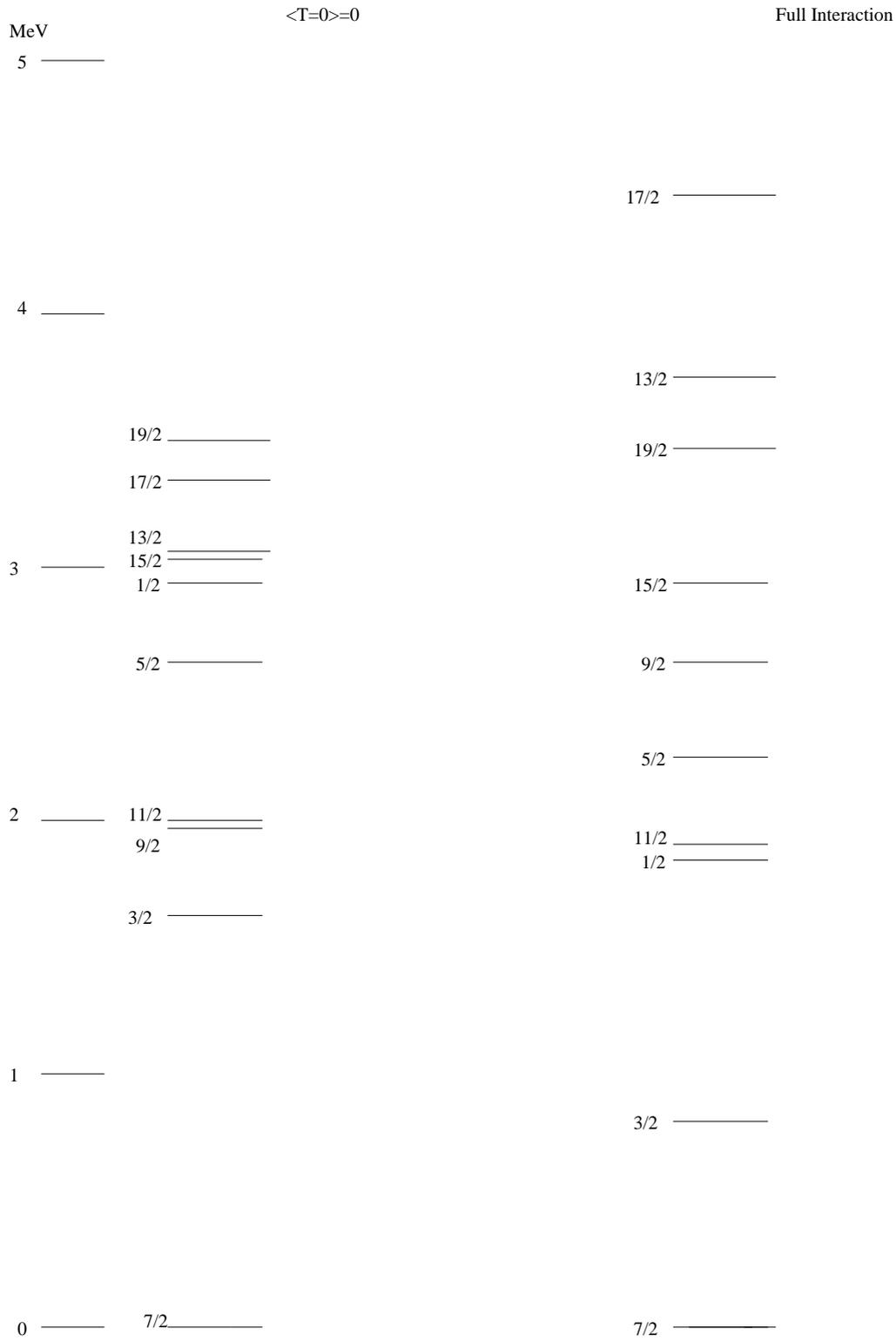,height=8in}
\caption{Full f-p $^{43}$Ti ($^{43}$Sc) with FPD6 interaction}
\end{figure}

\newpage

\centerline{References}

\begin{enumerate}
\item B.F. Bayman, J.D. McCullen and L. Zamick, Phys Rev Lett
\underline{10}, (1963) 117
 
\item J.D. McCullen, B.F. Bayman, and L. Zamick, Phys Rev
\underline{134}, (1964) 515; ``Wave Functions in the 1f$_{7/2}$
Shell'', Technical Report NYO-9891

\item J.N. Ginocchio and J.B. French Phys. Lett \underline{7}, (1963)
137

\item W. Kutschera, B.A. Brown and K. Ogawa Riv. Nuovo Cimento
\underline{1}, (1978) 12

\item J.A. Sheikh and R. Wyss, Phys. Rev \underline{C62}, (2000) 051302(R)

\item A. Goodman, Proceedings of the International Workshop on N=Z nuclei, Pinget 2000, Land

\item K. Muhlhass, E.M. Muller, K. Neergard and U. Mosel Phys. Lett. 105B (1981) 329

\item J. Terasaki, R Wyss, and P.-H Heenen, Phys. Lett. B437 (1998) 1

\item W. Satula and R. Wyss, Nucl. Phys \underline{A676}, (2000) 120

\item G. de Angelis et. al. Phys Lett \underline{B415}, (1997) 217

\item A.L. Goodman, Nucl Phys. \underline{A186}, (1972) 475

\item A.L. Goodman Phys Rev \underline{C60}, (1999) 014311 and references therein

\item A.O. Macchiavelli et. al. Phys. Lett \underline{B480} (2000) 1;
Los Alamos e-preprint Nucl 9805015

\item W.A. Richter, M.G. Van der Merwe, R.E. Julies and B.A. Brown,
Nucl. Phys. \underline{A253}, (1991) 325

\item P. Goode and L. Zamick, Phys. Lett. \underline{B43}, (1973)183

\item Quantum Theory of Angular Momentum, L.C. Biedenharn and H. Van Dam, Academic Press New York (1965)

\item T. Regge, Il Nuovo Cimento, Vol XI, N.1 (1959) 298

\item The 3-j and 6-j symbols, M. Rotenberg, R. Bivins, N. Metropolis,
and J.K. Wooten Jr, Technology Press M.I.T., Cambridge , MA (1959)

\item L.C. Biedenharn, J.M. Blatt, and M.E. Rose, Review of Modern Physics, \underline{24}, \#4, (1952) 212

\item V. Bargmann, Review of Modern Physics, \underline{34}, (1962) 300

\item Quantum Theory of Angular Momentum, D.A. Varshalovich, A.N. Moskalev, and V.K. Khersonski, World Scientific, Sinapore (1988)

\item A. Poves and G. Martinez-Pinedo Phys Lett. \underline{B430} (1998)203
\end{enumerate}

\end{document}